\documentclass[aps,pra,reprint,groupedaddress]{revtex4-2}
\usepackage{graphicx}
\usepackage{epstopdf}
\graphicspath{figures/}
\usepackage{dcolumn}
\usepackage{amssymb}
\usepackage{amsmath}
\usepackage{bm}
\usepackage{times}
\usepackage{mhchem}
\usepackage{lineno}
\usepackage[colorlinks,linkcolor=blue,anchorcolor=blue,citecolor=blue,urlcolor=blue]{hyperref}
\UseRawInputEncoding
\begin{document}
\title{Skyrmion-mechanical hybrid quantum systems: Manipulation of skyrmion qubits via phonons}
\author{Xue-Feng Pan}
\author{Xin-Lei Hei}
\author{Xiao-Yu Yao}
\author{Jia-Qiang Chen}
\author{Yu-Meng Ren}
\author{Xing-Liang Dong}
\author{Yi-Fan Qiao}
\affiliation{Ministry of Education Key Laboratory for Nonequilibrium Synthesis and Modulation of Condensed Matter, Shaanxi Province Key Laboratory of Quantum Information and Quantum Optoelectronic Devices, School of Physics, Xi'an Jiaotong University, Xi'an 710049, China}
\author{Peng-Bo Li}
\email{lipengbo@mail.xjtu.edu.cn}
\affiliation{Ministry of Education Key Laboratory for Nonequilibrium Synthesis and Modulation of Condensed Matter, Shaanxi Province Key Laboratory of Quantum Information and Quantum Optoelectronic Devices, School of Physics, Xi'an Jiaotong University, Xi'an 710049, China}

\date{\today}
\begin{abstract}
Skyrmion qubits are a new highly promising logic element for quantum information processing. However, their scalability to
multiple interacting qubits remains challenging. We propose a hybrid quantum setup with skyrmion qubits strongly coupled to nanomechanical cantilevers via magnetic coupling, which harnesses phonons as quantum interfaces for the manipulation of distant skyrmion qubits. A linear drive is utilized to achieve the modulation of the stiffness coefficient of the cantilever, resulting in an exponential enhancement of the coupling strength between the skyrmion qubit and the mechanical mode. We also consider the case of a topological resonator array,  which allows us to study interactions between skyrmion qubits and topological phonon band structure, as well as chiral skyrmion-skyrmion interactions. The scheme suggested here offers a fascinating platform for investigating quantum information processing and quantum simulation with magnetic microstructures.
\end{abstract}

\maketitle

\section{\label{sec:I}Introduction}
With the advancement of experimental techniques and theories, qubits, the core of quantum computation and quantum information processing, are becoming increasingly abundant and better performing~\cite{2013XiangP623653,2020ClerkP257267,2021BlaisP2500525005,2022ShandilyaP75387571}. In addition to the well-known superconducting qubits~\cite{1996ZorinP44084411,1999NakamuraP786788,1999MooijP10361039,2010MarcosP210501210501,2022ChirolliP177701177701,2023SomoroffP267001267001}, trapped ions or atoms~\cite{2011BulutaP104401104401}, and solid-state spins~\cite{2013MacQuarrieP227602227602,2013DohertyP145,2013BarGillP17431743,2014HeppP3640536405,2014TeissierP2050320503,2016GolterP143602143602,2018LemondeP213603213603,2018BeckerP5360353603,2019BradacP56255625,2023OrphalKobinP1104211042}, soliton-based qubits in magnetic insulators, such as two-dimensional skyrmions~\cite{2017FertP1703117031,2006RoesslerP797801,2009MuehlbauerP915919,2010YuP901904,2018EverschorSitteP240901240901,2019OchoaP19300051930005,2020LonskyP100903100903,2020BackP363001363001,2022ReichhardtP3500535005}, one-dimensional magnetic domain walls~\cite{2005KlaeuiP2660126601,2008ParkinP190194,2011MironP419423,2013EmoriP611616,2013RyuP527533}, and two-dimensional magnetic merons~\cite{2007HertelP117201117201,2008CurcicP197204197204,2012ImP983983,2013WintzP177201177201,2019GaoP56035603,2021WangP104408104408,2021AugustinP185185}, have recently received much attention due to their small lateral size, high topological robustness, and convenience of modulation~\cite{2012MochizukiP1760117601,2013LinP6040460404,2014SchuetteP9442394423,2015RoldanMolinaP245436245436,2015ZhangP102401102401,2017PsaroudakiP4104541045,2018PsaroudakiP237203237203,2019CasiraghiP145145,2020CapicP415803415803,2020HirosawaP207204207204,2021KhanP100402100402,2021LiensbergerP100415100415,2022HirosawaP4032140321}. Magnetic merons encode quantum information in chirality and polarity~\cite{2022XiaP8888}, whereas magnetic domain walls and skyrmions encode quantum information in collective coordinates~\cite{2018TakeiP6440164401,2023LiP23087515,2023ZouP3316633166,2021PsaroudakiP6720167201,2023XiaP106701106701}.  Skyrmions in frustrated magnetic materials, in particular, have an internal degree-of-freedom helicity $\varphi_0$ (collective coordinates)~\cite{2015LeonovP82758275,2016LinP6443064430,2022PsaroudakiP104422104422,2021PsaroudakiP6720167201,2023XiaP106701106701} that can be modulated by an electric field~\cite{2020YaoP8303283032} and can be utilized to construct a qubit by quantizing $\varphi_0$~\cite{2021PsaroudakiP6720167201,2023XiaP106701106701}. The skyrmion qubit has also been utilized to build universal qubit gates, which have emerged as potential candidates for quantum computer implementation~\cite{2023XiaP106701106701}. However, how to implement long-range interactions between distant skyrmion qubits remains unresolved.

The scalability of qubits and long-range interactions between them are essential for achieving quantum computation, which can be accomplished through indirect coupling mediated by a third quantum system like nanomechanical resonators~\cite{2013XiangP623653}.
With the development of manufacturing technology and nanofabrication, the lifetime and quality factor of the vibrational modes of nanomechanical resonators are constantly improving~\cite{2012PootP273335,2014AspelmeyerP13911452,2017DegenP3500235002,2020XiaP,2020WangP,2021PerdriatP651651,2022BarzanjehP1524,2022BachtoldP4500545005}. Commonly used nanomechanical resonators  include nanomechanical cantilevers~\cite{2009RablP4130241302,2010RablP602608,2011ArcizetP879883,2012KolkowitzP16031606,2009XuP2233522335,2010ZhouP4232342323,2013ChotorlishviliP8520185201}, clamped beams~\cite{2013BennettP156402156402,2013KepesidisP6410564105,2015LiP4400344003,2021ChenP1370913709}, ring oscillators~\cite{2021ShandilyaP14201425,2022YaoP5400454004},  carbon nanotubes~\cite{2016LiP1550215502,2012PootP273335}, optomechanical crystals~\cite{2009EichenfieldP7882,2011ChanP8992,2013SafaviNaeiniP185189,2014SafaviNaeiniP153603153603,2017FangP465471,2018RiedingerP473477,2021DongP203601203601}, suspended particles~\cite{2015MillenP123602123602,2020DelicP892895,2020TebbenjohannsP1360313603,2020GieselerP163604163604,2021StreltsovP193602193602,2021GonzalezBallesteroP30273027,2022RusconiP9360593605,2023PanP2372223722}, and so on; they can couple to other quantum systems such as solid-state spins~\cite{2009RablP4130241302,2009XuP2233522335,2010RablP602608,2010ZhouP4232342323,2011ArcizetP879883,2012KolkowitzP16031606,2013ChotorlishviliP8520185201,2013BennettP156402156402,2013KepesidisP6410564105,2015LiP4400344003,2021ChenP1370913709,2016LiP1550215502,2021ShenP21000742100074,2022QiaoP3241532415}, quantum dots~\cite{2018CarterP246801246801}, and magnons~\cite{2018LiP203601203601,2022ShenP243601243601} through magnetic dipole interactions~\cite{2009RablP4130241302,2010RablP602608,2011ArcizetP879883}, strain coupling~\cite{2013MacQuarrieP227602227602,2014OvartchaiyapongP44294429,2013BennettP156402156402}, and magnetostrictive effects~\cite{2018LiP203601203601,2022ShenP243601243601}. Mechanical oscillator-based hybrid quantum systems have been widely employed in the research of quantum effects such as entanglement~\cite{2013ChotorlishviliP8520185201,2023HeiP7360273602,2018RiedingerP473477}, quantum transduction~\cite{2010RablP602608,2010ZhouP4232342323}, ground state cooling~\cite{2011ChanP8992,2013KepesidisP6410564105,2021StreltsovP193602193602}, squeezed states~\cite{2013BennettP156402156402,2021ChenP1370913709,2013SafaviNaeiniP185189}, and nonreciprocal phenomena~\cite{2022YaoP5400454004,2017FangP465471}. Furthermore, a parametric amplification technique based on the nanomechanical system has been proposed and demonstrated~\cite{2020LiP153602153602,2021BurdP898902}, which is important for enhancing the coupling strength between mechanical resonators and other quantum systems.

In this work, we propose a hybrid quantum system with a skyrmion qubit strongly coupled to a nanomechanical cantilever via a magnetic gradient field generated by a magnetic tip. We show that  coherent coupling between phonons and skyrmion qubits can reach the strong coupling regime, indicating that the phonons can be employed as a quantum interface for manipulating the skyrmion qubits. Mechanical amplification can be realized experimentally by positioning an electrode near the lower surface of the cantilever and applying a tunable and time-varying voltage to this electrode. This effect leads to a two-phonon drive and consequently an exponential increase in the phonon-skyrmion coupling strength~\cite{2020LiP153602153602,2021BurdP898902}. We then investigate coherent interactions between distant skyrmion qubits via a single mechanical resonator, showing that strong skyrmion-skyrmion coupling is possible in the presence of a parametric drive. Furthermore, we extend a single cantilever to a coupled resonator array  with a topological phonon structure. When a single skyrmion qubit is coupled to the topological phononic bath, a chiral skyrmion-phonon bound state can be obtained due to the topological characteristics of the phononic bath, and the chirality can be controlled by adjusting the two-phonon drive. It can be further used to mediate chiral skyrmion-skyrmion interactions. The modulation of the chiral coupling can be achieved by adjusting the position of the skyrmion qubit and the phonon hopping rate.

\section{\label{sec:II}The setup}
\subsection{\label{sec:IIA}The Hamiltonian of skyrmion qubits and phonons}
\begin{figure}
    \centering
    \includegraphics[width=0.36\textwidth]{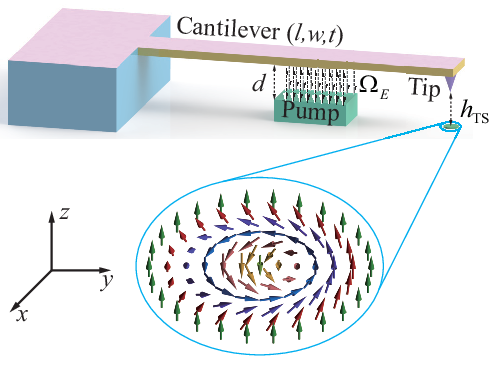}
    \caption{The model of our proposal. The geometry of the cantilever is $(l,w,t)$, and the distance from the driving electrode to the cantilever is $d$. The insert depicts the structure of a skyrmion.}
    \label{Fig1}
\end{figure}

As illustrated in Fig.~\ref{Fig1}, a skyrmion in frustrated magnets is a spin texture with a centrosymmetric vortex structure that is supposed to lie in the $xy$ plane and holds its center at the origin of the coordinates. The Hamiltonian of the skyrmion in frustrated magnets can be given by~\cite{2021PsaroudakiP6720167201,2016LinP6443064430}
\begin{equation}
    \mathcal{H}=\int d\widetilde{\boldsymbol{r}}\left[-\frac{\mathcal{J}_1}{2}(\nabla_{\widetilde{\boldsymbol{r}}}\boldsymbol{s})^2+\frac{\mathcal{J}_2 a^2 }{2}(\nabla_{\widetilde{\boldsymbol{r}}}^2\boldsymbol{s})^2-\frac{H}{a^2}s_z+\frac{K}{a^2}s_z^2\right]
    \label{GLEF}
\end{equation}
with competing interactions $\mathcal{J}_{1/2}$, lattice spacing $a$, position vector $\widetilde{\boldsymbol{r}}=(\widetilde{\rho},\phi)$, $z$-direction magnetic field $H$, and $z$-direction easy-axis anisotropy $K$, according to Ginzburg-Landau theory~\cite{2016LinP6443064430}.  There is $\boldsymbol{s}=[\sin \Theta(\widetilde{\rho}) \cos \Phi,\sin \Theta(\widetilde{\rho}) \sin \Phi,\cos \Theta(\widetilde{\rho})]$ for a classical skyrmion with $\Phi=\phi+\varphi_0$ and helicity $\varphi_0$. We can derive the steady-state solution, represented by $\Phi_0$ and $\Theta_0$, by minimizing the energy $\mathcal{H}$ (Appendix~\ref{sec:appendixA}).

Skyrmion qubits can be designed by quantizing the helicity $\varphi_0$~\cite{2021PsaroudakiP6720167201}. The Hamiltonian of $\mathfrak{S}_z$ qubits, employing the collective coordinate quantization technique~\cite{1975GoldstoneP14861498,1994DoreyP35983611,2021PsaroudakiP6720167201,2022PsaroudakiP104422104422}, is given by
\begin{equation}\label{z}
\hat{\mathcal{H}}_{\mathfrak{S}_z}=\bar{\bar{\kappa}}_z\hat{\mathfrak{S}}_z^2-\bar{\bar{h}}_z\hat{\mathfrak{S}}_z-\bar{\bar{\varepsilon}}_z\cos\hat{\varphi}_0,
\end{equation}
with the collective coordinate $\hat{\varphi}_0$ and its conjugate momentum $\hat{\mathfrak{S}}_z$, where the parameters $\bar{\bar{\kappa}}_z$, $\bar{\bar{h}}_z$, and $\bar{\bar{\varepsilon}}_z$ are described in detail in appendix~\ref{sec:appendixB}. Because the energy level of $\mathfrak{S}_z$ qubits is non-uniform (the non-uniformity satisfies $\vert\omega_{ex}-\omega_q\vert>20\%\omega_q$, where $\omega_q$ is the qubit frequency defined below and $\omega_{ex}$ is the frequency of higher level transitions~\cite{2021PsaroudakiP6720167201, 2023PsaroudakiP260501260501}), we can truncate the Hilbert space to $\vert 0\rangle$ and $\vert 1\rangle$, simplifying the Hamiltonian of $\mathfrak{S}_z$ qubits to
\begin{equation}
    \hat{H}_{\rm{Sky}}=\frac{\mathcal{A}_0}{2}\hat{\sigma}_z^{\rm{sub}}-\frac{\mathcal{B}_0}{2}\hat{\sigma}_x^{\rm{sub}}
    \label{HSky}
\end{equation}
with $\mathcal{A}_0\equiv\bar{\bar{\kappa}}_z-\bar{\bar{h}}_z$, $\mathcal{B}_0\equiv\bar{\bar{\varepsilon}}_z$, and Pauli operators $\hat{\sigma}_z^{\rm{sub}}\equiv\vert1\rangle\langle1\vert-\vert0\rangle\langle0\vert$ and $\hat{\sigma}_x^{\rm{sub}}\equiv\vert1\rangle\langle0\vert+\vert0\rangle\langle1\vert$.

The cantilever with dimensions $(l,w,t)$ depicted in Fig.~\ref{Fig1}, can be described by the Hamiltonian
\begin{equation}
    \hat{H}_m=\frac{\hat{p}^2}{2m}+\frac{1}{2}k_0\hat{z}^2,
    \label{Hm}
\end{equation}
where $\hat{p}_z$ and $\hat{z}$ are momentum and position operators, respectively. With $\hat{p}_z=-i\sqrt{m\omega_{m}/2} (\hat{a}-\hat{a}^{\dagger})$, $\hat{z}=z_0 (\hat{a} + \hat{a}^{\dagger})$, and $z_0=\sqrt{1/ (2m \omega_{m})}$, the Hamiltonian can be written as (setting $\hslash=1$)
\begin{equation}
    \hat{H}_{m} = \omega_{m} \hat{a}^{\dagger} \hat{a},
    \label{HmQ}
\end{equation}
where $\omega_{ma}=\sqrt{k_{ma}/m}$ indicates the resonance frequency and $\hat{a}$ ($\hat{a}^{\dagger}$) represents the annihilation (creation) operator for the phonon mode.
\subsection{\label{sec:IIB}Interaction Hamiltonian}
\begin{figure}
    \centering
    \includegraphics[width=0.48\textwidth]{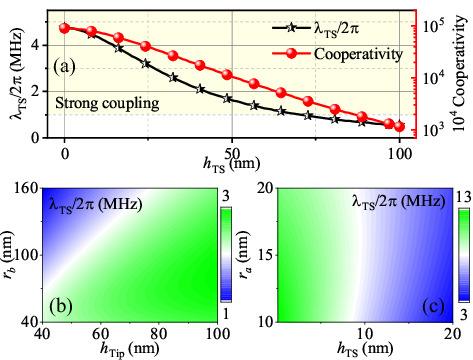}
    \caption{(a) The cooperativity $\mathcal{C}_{\rm{TS}}=4\lambda_{\rm{TS}}^2/(\gamma_m\gamma_{\rm Sky})$ and coupling strength $\lambda_{\rm{TS}}$ are displayed as a function of the distance $h_{\rm{TS}}$ from the magnetic tip to the skyrmion qubit. Shaded areas indicate strong coupling regions $\lambda_{\rm{TS}}>\max \{\gamma_{\rm Sky},\gamma_m\}$. The dissipation rates of skyrmions and phonons used in the calculation of the cooperativity is $\gamma_m/2\pi=1~{\rm kHz}$ and $\gamma_{\rm{Sky}}/2\pi=1~{\rm{MHz}}$, respectively. (b) and (c) depict the relationship between the coupling strength and magnetic tip shape. The parameters $h_{\rm{Tip}}$, $r_a$, and $r_b$ represent the height of the magnetic tip, the radius of the upper bottom surface, and the radius of the lower bottom surface, respectively.}
    \label{Fig2}
\end{figure}

The mechanical mode and the skyrmion qubit are coupled by the magnetic gradient field generated by the magnetic tip, which can be described by the Hamiltonian~\cite{2021PsaroudakiP6720167201}
\begin{equation}
    \hat{H}_{\rm{TS}}=-\frac{g\mu_B \bar{S}}{a^2}\int d\widetilde{\boldsymbol{r}} \boldsymbol{B}\cdot \boldsymbol{s}
    \label{IHOrigin}
\end{equation}
with the Land\'e g-factor $g$, the Bohr magneton $\mu_B$, and the effective spin $\bar{S}$. Substituting the gradient magnetic fields $\boldsymbol{B}_z=-G\widetilde{z}\boldsymbol{e}_z$~(see Appendix~\ref{sec:appendixC}) and $\boldsymbol{s}$ into Eq.~(\ref{IHOrigin}) and dimensionless the integral results in
\begin{equation}
    \hat{H}_{\rm{TS}}=g\mu_B\bar{S}G\widetilde{z}\int dxdy\left(1-\widetilde{\Pi}\right),
    \label{IHI}
\end{equation}
where $\widetilde{\Pi}\equiv 1-\Pi$ and $\Pi\equiv \cos\Theta$. \textcolor[rgb]{0,0,0}{Here we have ignored in-plane component of the tip's magnetic field}, since it is much smaller than the magnetic field in the perpendicular direction produced by the tip~\cite{2009RablP4130241302,2010RablP602608,2020LiP153602153602,2011ArcizetP879883,2012KolkowitzP16031606,2009DegenP13131317}. Using Eq.~(\ref{CCQ}), the interaction Hamiltonian can be reduced to
\begin{equation}
    \hat{H}_{\rm{TS}}=-\lambda_{\rm{TS}}\left(\hat{a}+\hat{a}^{\dagger}\right)\hat{\mathfrak{S}}_z,
\end{equation}
where $\lambda_{\rm{TS}}=g\mu_B\bar{S}Gz_0$ is the coupling strength. Figure~\ref{Fig2} depicts the phonon-skyrmion coupling strength versus the distance between the magnetic tip and the skyrmion, as well as the magnetic tip geometry. The coupling strength decreases with increasing distance $h_{\rm{TS}}$, as illustrated in Fig.~\ref{Fig2}(a). It should be noted that this hybrid quantum system is capable of entering the strong-coupling region [the shaded region in Fig.~\ref{Fig2}(a)]. Figures~\ref{Fig2}(b, c) demonstrate that the coupling strength increases as the magnetic tip becomes sharper (smaller $r_a$).
In the subspace $\{\vert 0\rangle,\vert 1\rangle\}$, the interaction Hamiltonian can be simplified to
\begin{equation}
    \hat{H}_{\rm{TS}}=-\frac{\lambda_{\rm{TS}}}{2}\left(\hat{a}+\hat{a}^{\dagger}\right)\hat{\sigma}_z^{\rm{sub}},
    \label{HIntOrigin}
\end{equation}
where $\hat{\sigma}_z^{\rm{sub}}\equiv \vert 1\rangle\langle 1\vert - \vert 0\rangle\langle 0\vert$.

Since the resonance frequency of the skyrmion qubit (GHz) is significantly greater than that of the mechanical mode (MHz), we employ the dressed state technique to establish resonance here. In other words, we realize the resonance between the mechanical oscillator and the qubit by applying microwave drive to the qubit and converting it to the dressed state basis~\cite{2009RablP4130241302,2020LiP153602153602}. The microwave-driven Hamiltonian is computed in the same method as the interaction Hamiltonian from Eq.~(\ref{IHOrigin}) to Eq.~(\ref{HIntOrigin}). With the microwave drive $\boldsymbol{B}_{\rm{mw}}=B_0\cos \left(\omega_{\rm{mw}}t\right)\boldsymbol{e}_z$ polarized in the $z$ direction, the Hamiltonian of the microwave drive can be expressed as
\begin{equation}
    \hat{H}_{\rm{mw}}=\Omega_{\rm{mw}}\cos \left(\omega_{\rm{mw}}t\right)\hat{\sigma}_z^{sub},
\end{equation}
where $B_0$ and $\omega_{\rm{mw}}$ are the magnetic field amplitude and frequency of the microwave drive, respectively. $\Omega_{\rm{mw}}\equiv g\mu_B B_0\bar{S}/2$ denotes the microwave-drive strength.

\subsection{\label{sec:IID}The phonon-skyrmion hybrid quantum system}
The total Hamiltonian of the hybrid quantum system is represented as
\begin{equation}
    \hat{H}_{\rm{TTS}}=\hat{H}_{\rm{Sky}}+\hat{H}_m+\hat{H}_{\rm{TS}}+\hat{H}_{\rm{mw}}.
\end{equation}
In this case, the Hamiltonian $\hat{H}_{\rm{Sky}}$ is not diagonalized. The eigenstates of the Hamiltonian $\hat{H}_{\rm{Sky}}$ are $\vert \psi_+ \rangle=\cos\theta\vert 1\rangle - \sin\theta \vert 0\rangle$ and $\vert \psi_- \rangle=\sin\theta\vert 1\rangle + \cos\theta \vert 0\rangle$, and the corresponding eigenenergies are $\mathcal{E}_{\pm}=\pm\sqrt{\mathcal{A}_0^2+\mathcal{B}_0^2}/2$ with $\tan(2\theta)\equiv \mathcal{B}_0/\mathcal{A}_0$. In terms of $\vert \psi_+ \rangle$ and $\vert \psi_- \rangle$, the Hamiltonian $\hat{H}_{\rm{Sky}}$ can be expressed as $\hat{H}_{\rm{Sky}}=\omega_q \hat{\sigma}_z^D/2$, where the resonance frequency of the skyrmion qubit is denoted by $\omega_q=\mathcal{E}_+-\mathcal{E}_-$. In the subspace $\left\{\vert \psi_+ \rangle,\vert \psi_- \rangle\right\}$, the interaction Hamiltonian $\hat{H}_{\rm{TS}}$ and the microwave-driven Hamiltonian $\hat{H}_{\rm{mw}}$ reduce to
\begin{subequations}
    \begin{equation}
        \hat{H}_{\rm{TS}}=-\frac{\lambda_{\rm{TS}}}{2}\left(\hat{a}+\hat{a}^{\dagger}\right)\left[\cos(2\theta)\hat{\sigma}_z^D+\sin(2\theta)\hat{\sigma}_x^D\right],
        \label{HIntOriginFrame}
    \end{equation}
    \begin{equation}
        \hat{H}_{\rm{mw}}=\Omega_{\rm{mw}}\cos\left(\omega_{\rm{mw}}t\right)\left[\cos(2\theta)\hat{\sigma}_z^D+\sin(2\theta)\hat{\sigma}_x^D\right]
        \label{HmwOriginFrame}
    \end{equation}
\end{subequations}
with Pauli operators $\hat{\sigma}_z^D=\vert \psi_+\rangle\langle \psi_+\vert-\vert \psi_-\rangle\langle \psi_-\vert$ and $\hat{\sigma}_x^D=\vert \psi_+\rangle\langle \psi_-\vert+\vert \psi_-\rangle\langle \psi_+\vert$. If the $\mathfrak{S}_z$ qubit is supposed to work near the degeneracy point, subsequently $\cos(2\theta)\sim 0$ and $\sin(2\theta)\sim 1$ can be obtained. The term containing $\cos(2\theta)$ can therefore be removed, reducing the Hamiltonian (\ref{HIntOriginFrame}) and (\ref{HmwOriginFrame}) to
\begin{subequations}
    \begin{equation}
        \hat{H}_{\rm{TS}}=-\frac{\lambda_{\rm{TS}}}{2}\left(\hat{a}+\hat{a}^{\dagger}\right)\hat{\sigma}_x^D,
        \label{HIntOriginFrameD}
    \end{equation}
    \begin{equation}
        \hat{H}_{\rm{mw}}=\Omega_{\rm{mw}}\cos\left(\omega_{\rm{mw}}t\right)\hat{\sigma}_x^D.
        \label{HmwOriginFrameD}
    \end{equation}
\end{subequations}

The Hamiltonian of the $\mathfrak{S}_z$ qubit
\begin{equation}
    \hat{H}_{\rm{dress}}=\frac{\omega_q}{2}\hat{\sigma}_z^D+\Omega_{\rm{mw}}\cos\left(\omega_{\rm{mw}}t\right)\hat{\sigma}_x^D
    \label{HDress}
\end{equation}
can be diagonalized by the eigenstates $\vert \widetilde{\psi}_+\rangle=\cos\beta\vert\psi_+\rangle+\sin\beta\vert\psi_-\rangle$ and $\vert \widetilde{\psi}_-\rangle=\sin\beta\vert\psi_+\rangle-\cos\beta\vert\psi_-\rangle$, with the corresponding eigenenergies $\widetilde{\mathcal{E}}_{\pm}=\pm 1/2 \sqrt{\Delta_{\rm qmw}^2 + \Omega_{\rm{mw}}^2}$. Here $\Delta_{\rm qmw}=\omega_q-\omega_{\rm{mw}}$ and $\tan(2\beta)=\Omega_{\rm{mw}}/\Delta_{\rm qmw}$. In this case, the total Hamiltonian of the hybrid system is
\begin{equation}
    \begin{split}
        \hat{H}_{\rm{TTS}}&=\frac{\widetilde{\omega}_q}{2}\hat{\sigma}_z+\omega_{m}\hat{a}^{\dagger}\hat{a}\\ &-\frac{\lambda_{\rm{TS}}}{2}\left(\hat{a}+\hat{a}^{\dagger}\right)\left[\sin(2\beta)\hat{\sigma}_z-\cos(2\beta)\hat{\sigma}_x\right],
    \end{split}
    \label{HTTSMW}
\end{equation}
where $\widetilde{\omega}_q\equiv\widetilde{\mathcal{E}}_+-\widetilde{\mathcal{E}}_-$, $\hat{\sigma}_z\equiv\vert\widetilde{\psi}_+\rangle\langle \widetilde{\psi}_+\vert-\vert\widetilde{\psi}_-\rangle\langle \widetilde{\psi}_-\vert$, and $\hat{\sigma}_x\equiv\vert\widetilde{\psi}_+\rangle\langle \widetilde{\psi}_-\vert+\vert\widetilde{\psi}_-\rangle\langle \widetilde{\psi}_+\vert$. We can attain $\sin(2\beta)\ll \cos(2\beta)$ by modifying the microwave drive, and then Hamiltonian~(\ref{HTTSMW}) can be represented as
\begin{equation}
    \hat{H}_{\rm{TTS}}=\frac{\widetilde{\omega}_q}{2}\hat{\sigma}_z+\omega_{m}\hat{a}^{\dagger}\hat{a}+\bar{\lambda}_{\rm{TS}}\left(\hat{a}+\hat{a}^{\dagger}\right)\hat{\sigma}_x,
    \label{HTTS}
\end{equation}
where the coupling strength is defined as $\bar{\lambda}_{\rm{TS}}\equiv \lambda_{\rm{TS}}\cos(2\beta)$.

\section{\label{sec:III}Modulation of the skyrmion qubits by a single mechanical oscillator}
\subsection{\label{sec:IIIA}Modulation of the phonon-skyrmion coupling strength}
\begin{figure}
    \centering
    \includegraphics[width=0.48\textwidth]{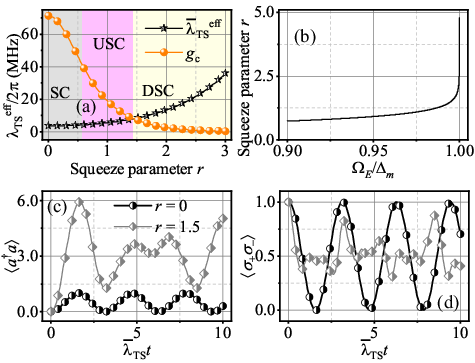}
    \caption{(a) The effective coupling strength $\lambda_{\rm{TS}}^{\rm{eff}}$ versus the squeezed parameter $r$. (b) displays the squeezed parameter $r$ as a function of the ratio $\Omega_E/\Delta_m$. (c) and (d) depict the influence of a two-phonon drive on the dynamics of a hybrid quantum system. The skyrmion occupation is defined as $\langle \hat{\sigma}_+\hat{\sigma}_-\rangle$. The other parameters are $\Delta_q=\Delta_m^{\rm eff}$, $\Delta_m^{\rm eff}=\Delta_m/\cosh(2r)$, $\Delta_m=20\bar{\lambda}_{\rm{TS}}$, $\gamma_{\rm{Sky}}/2\pi=0$, and $\gamma_m/2\pi=0$.}
    \label{Fig3}
\end{figure}
To manipulate the coupling strength of the system, a voltage drive is used to alter the stiffness coefficient of the mechanical oscillator~\cite{2020LiP153602153602,2021BurdP898902}. The hybrid system with the mechanical oscillator driven by a voltage can be described by the Hamiltonian (see Appendix~\ref{sec:appendixD})
\begin{equation}
    \begin{split}
        \hat{H}_{\rm{TTS}}&=\frac{\Delta_q}{2}\hat{\sigma}_z+\Delta_{m}\hat{a}^{\dagger}\hat{a}+\bar{\lambda}_{\rm{TS}}\left(\hat{a}\hat{\sigma}_++\hat{a}^{\dagger}\hat{\sigma}_-\right) \\&-\frac{\Omega_E}{2}\left({\hat{a}^{\dagger}}{^2}+\hat{a}^2\right)
        \label{HTTSDD}
    \end{split}
\end{equation}
with the detunings defined as $\Delta_q\equiv\widetilde{\omega}_q-\omega_E$ and $\Delta_m\equiv\omega_m-\omega_E$. Employing the Bogoliubov transformation $\hat{b}=\hat{a}\cosh r-\hat{a}^{\dagger}\sinh r$~\cite{2016LemondeP1133811338,2019BurdP11631165,2021BurdP898902}, we transform the Hamiltonian ~(\ref{HTTSDD}) into the squeezed picture resulting in
\begin{equation}
    \hat{H}_{\rm{TTS}}^{\rm{sq}}=\frac{\Delta_q}{2}\hat{\sigma}_z+\Delta_m^{\rm{eff}}\hat{b}^{\dagger}\hat{b}+\lambda_{\rm{TS}}^{\rm{eff}}\left(\hat{b}+\hat{b}^{\dagger}\right)\hat{\sigma}_x
    \label{HTTSSqr}
\end{equation}
with $\Delta_m^{\rm{eff}}=\Delta_m/\cosh(2r)$. The phonon-skyrmion coupling strength $\lambda_{\rm{TS}}^{\rm{eff}}=\bar{\lambda}_{\rm{TS}}\exp(r)/2$ is increased exponentially [Fig.~\ref{Fig3}(a)]. The squeezed parameter $r$ is related to the ratio of the voltage drive $\Omega_E$ and detuning $\Delta_m$, which is defined as $\tanh (2r)=\Omega_E/\Delta_m$ [Fig.~\ref{Fig3}(b)]. We introduce the parameter $g_c=\sqrt{\Delta_q\Delta_m^{\rm{eff}}}$ to characterize the different coupling regions. Considering the near-resonance condition $\Delta_q\approx\Delta_m^{\rm{eff}}$, it reduces to $g_c\approx\Delta_m^{\rm{eff}}$. The region of ultrastrong coupling (USC) is defined as $0.1\lesssim\lambda_{\rm{TS}}^{\rm{eff}}/g_c=\lambda_{\rm{TS}}^{\rm{eff}}/\Delta_m^{\rm{eff}}\lesssim 1$, and the region of deep strong coupling (DSC) is localized as $\lambda_{\rm{TS}}^{\rm{eff}}/g_c=\lambda_{\rm{TS}}^{\rm{eff}}/\Delta_m^{\rm{eff}}\gtrsim 1$~\cite{2019FornDiazP2500525005}. As shown in Fig.~\ref{Fig3}(a), as the squeezing parameter $r$ increases, the system can gradually transition from the strong coupling (SC) region to the USC region, and finally can enter the DSC region. The dynamical evolution of the Hamiltonian~(\ref{HTTSSqr}) in the SC region and near the USC and DSC boundary is illustrated in Fig.~\ref{Fig3}(c, d), \textcolor[rgb]{0,0,0}{which is described by the Lindblad master equation
\begin{equation}
    \dot{\hat{\rho}}=-i[\hat{H}_{\rm{TTS}}^{\rm{sq}},\hat{\rho}]+\gamma_{m}D[\hat{b}]+\gamma_{\rm{Sky}}D[\hat{\sigma}_-]+\gamma_{\rm{Sky}}D[\hat{\sigma}_z]
\end{equation}
where $D[\hat{O}]\cdot=\hat{O}\cdot\hat{O}^\dagger-\{\hat{O}^\dagger\hat{O},\cdot\}/2$ is the Lindblad operator and $\{\gamma_m, \gamma_{\rm{Sky}}\}$ is the dissipation rate of the system. Here, the decay $D[\hat{b}]$ of the phonon, the decay $D[\hat{\sigma}_-]$ and the dephasing $D[\hat{\sigma}_z]$ of the skyrmion qubit are considered}. It can be seen that in the SC region ($r=0$), the Hamiltonian~(\ref{HTTSSqr}) can be approximated as a Jaynes-Cummings model and the dynamics of the system behaves as a standard Rabi oscillation. As the squeezing parameter $r$ increases and the system enters the USC or even DSC region ($r=1.5$), the Hamiltonian~(\ref{HTTSSqr}) will be described only by the Rabi model, in which case the occupation of the phonons can be larger than one.

\subsection{\label{sec:IIIB}Phonon-mediated effective skyrmion-skyrmion interactions}
\begin{figure}
    \centering
    \includegraphics[width=0.48\textwidth]{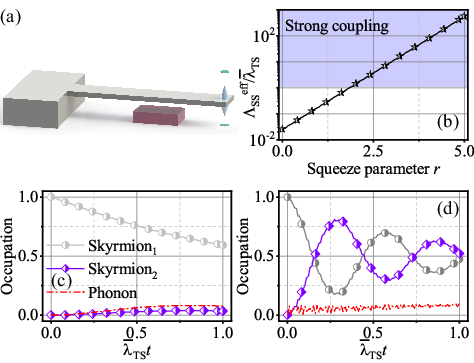}
    \caption{(a) A scheme of skyrmion-skyrmion interactions mediated by phonons. (b) As the squeezed parameter $r$ grows, the effective coupling strength $\Lambda_{\rm{SS}}^{\rm{eff}}$ increases exponentially. (c) and (d) depict direct state conversion between two skyrmion qubits in the absence ($r=0$) and presence ($r=4$) of the parametric drive, respectively. The parameters used here are $\gamma_{\rm{Sky}}=0.5\bar{\lambda}_{\rm{TS}}$, $\gamma_m=0.1\bar{\lambda}_{\rm{TS}}$, $\Delta_m^{\rm{eff}}=10\lambda_{\rm{TS}}^{\rm{eff}}$, and $\Delta_q=0$.}
    \label{Fig4}
\end{figure}
This section investigates skyrmion-skyrmion interactions mediated by a single mechanical oscillator. Two skyrmions depicted in Fig.~\ref{Fig4}(a) are situated on opposite sides of a cantilever. When a voltage drive is applied to the mechanical oscillator, the Hamiltonian of the hybrid quantum system is displayed as
\begin{equation}
    \begin{split}
        \hat{H}_{\rm{TSTS}}&=\frac{\widetilde{\omega}_q}{2}\left(\hat{\sigma}_z^1+\hat{\sigma}_z^2\right)+\omega_m\hat{a}^{\dagger}\hat{a} \\ &+\bar{\lambda}_{\rm{TS}}\left(\hat{a}+\hat{a}^{\dagger}\right)\left(\hat{\sigma}_x^1-\hat{\sigma}_x^2\right)\\
        &-\Omega_E\cos\left(2\omega_E t\right)\left(\hat{a}+\hat{a}^{\dagger}\right)^2.
    \end{split}
    \label{HTSTSOrigin}
\end{equation}
Transforming to the squeezed picture gives
\begin{equation}
    \begin{split}
        \hat{H}_{\rm{TSTS}}^{\rm{sq}}&=\frac{\Delta_q}{2}\left(\hat{\sigma}_z^1+\hat{\sigma}_z^2\right)+\Delta_m^{\rm{eff}}\hat{b}^{\dagger}\hat{b}\\ &+\lambda_{\rm{TS}}^{\rm{eff}}\left(\hat{b}+\hat{b}^{\dagger}\right)\left(\hat{\sigma}_x^1-\hat{\sigma}_x^2\right).
    \end{split}
    \label{HTSTSSqr}
\end{equation}
The foregoing analysis shows that the coupling strength $\lambda_{\rm{TS}}^{\rm{eff}}$ is enhanced exponentially. The effective skyrmion-skyrmion interaction can be achieved by adiabatically eliminating the phonon modes. Utilizing the Schrieffer-Wolff (SW) transformation $\hat{U}=e^{\epsilon\hat{S}}$ with $\epsilon=-i$, $\hat{S}=\hat{P}(\hat{\sigma}_x^1-\hat{\sigma}_x^2)$, and $\hat{P}=i\lambda_{\rm{TS}}^{\rm{eff}}/\Delta_m^{\rm{eff}}(\hat{b}^{\dagger}-\hat{b})$~\cite{2004WilsonRaeP7550775507,2013AlbrechtP8301483014}, the effective Hamiltonian of the hybrid quantum system is
\begin{equation}
    \hat{H}_{\rm{TSTS}}^{\rm{eff}}=\Lambda_{\rm{SS}}^{\rm{eff}}\left(\hat{\sigma}_x^1-\hat{\sigma}_x^2\right)^2,
\end{equation}
where the effective coupling strength can be calculated by $\Lambda_{\rm{SS}}^{\rm{eff}}=\lambda_{\rm{TS}}^{\rm{eff}}{^2}/\Delta_m^{\rm{eff}}$. Figure ~\ref{Fig4}(b) shows that when the squeezed parameter $r$ increases, the coupling strength increases exponentially and can approach the strong coupling region.\textcolor[rgb]{0,0,0}{The dynamical evolution is described by the Lindblad master equation
\begin{equation}
    \dot{\hat{\rho}}=-i[\hat{H}_{\rm{TSTS}}^{\rm{sq}},\hat{\rho}]+\gamma_{m}D[\hat{b}]+\gamma_{\rm{Sky}}D[\hat{\sigma}_-]+\gamma_{\rm{Sky}}D[\hat{\sigma}_z]
\end{equation}
where $D[\hat{O}]\cdot=\hat{O}\cdot\hat{O}^\dagger-\{\hat{O}^\dagger\hat{O},\cdot\}/2$ is the Lindblad operator}. As illustrated in Fig.~\ref{Fig4}(c), in the absence of the parametric drive, the effective skyrmion-skyrmion coupling strength is weak and there is no state conversion between the skyrmion qubits. When the parametric drive is applied, the skyrmion-skyrmion coupling increases exponentially, and state conversions between the skyrmions occur [Fig.~\ref{Fig4}(d)]. The phonon modes are adiabatically eliminated due to the large detuning condition, and hence their occupation is always $0$.

\section{\label{sec:IV}Modulation of the skyrmion qubits by a mechanical resonator array}
\subsection{\label{sec:IVA}The Hamiltonian of the hybrid system}
\begin{figure}
    \centering
    \includegraphics[width=0.48\textwidth]{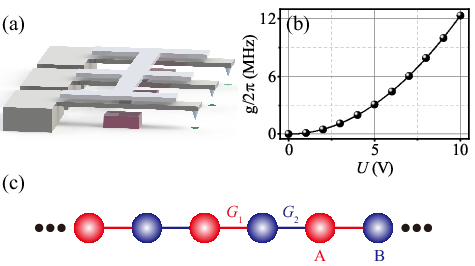}
    \caption{(a) Modulation of multi-skyrmion interactions utilizing an oscillator array. (b) The hopping rate between phonons can be controlled by the applied voltage $U$. The parameters utilized here are $d=100~\rm{\mu m}$, $h=100~\rm{nm}$, and $C=0.1~\rm{fF}$. (c) depicts the SSH model obtained from the mechanical oscillator array in (a).}
    \label{Fig5}
\end{figure}

In the preceding section, we explored the skyrmion-skyrmion interaction mediated by a single mechanical resonator. When we expand the single mechanical resonator into a resonator array, the dynamics of the hybrid quantum system become more enriched. The different resonators in the array can couple to each other by capacitances, as shown in Fig.~\ref{Fig5}(a), and the coupling strength can be calculated by computing the electrostatic energy of the circuit. The Hamiltonian of the coupling of skyrmion qubits and the oscillator array can be written as (see Appendix~\ref{sec:appendixE})
\begin{equation}
    \begin{split}
        \hat{H}_{\rm{MTMS}}^{\rm{sq}}&=\sum_n\frac{\Delta_q}{2}\hat{\sigma}_z^n+\sum_n\Omega_m\hat{b}_n^{\dagger}\hat{b}_n\\ &+\sum_n\mathcal{G}\left(\hat{b}_n\hat{\sigma}_+^n+H.c.\right)
        +\sum_n G_n\left(\hat{b}_n^{\dagger}\hat{b}_{n+1}+H.c.\right).
    \end{split}
    \label{HMTMSSqr}
\end{equation}
Here, we have assumed $\Delta_m^{\rm{eff}}\sim\Delta_q\gg G_n$. All the parameters in Eq.~(\ref{HMTMSSqr}) are defined as $\mathcal{G}\equiv\lambda_{\rm{TS}}^{\rm{eff}}=\bar{\lambda}_{\rm{TS}}\exp(r)/2$, $G_n\equiv g_n^{\rm{eff}}=g_n\exp(2r)/2$, and $\Omega_m\equiv \Delta_{m}^{\rm{eff}}$. It is worth noting that the bare hopping rate $g_n=z_0^2U_n^2C^2C_W^2/[h^2(2C+C_W)^3]$ [see Appendix~\ref{sec:appendixE}] can be modulated by varying the applied voltage $U$ [Fig.~\ref{Fig5}(b)].

\begin{figure*}
    \centering
    \includegraphics[width=0.65\textwidth]{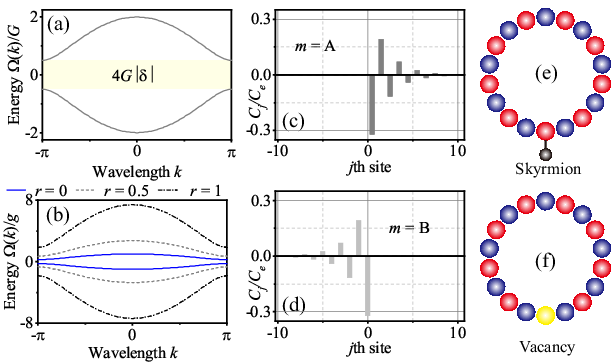}
    \caption{(a) Dispersion relations for the phononic bath with $\delta=0.25$. (b) The dispersion relation is affected by the two-phonon drive. Here, $\delta=0.25$ is used. (c, d) show chiral skyrmion-phonon bound states. The parameters are $E_{\rm{BS}}=0$, $\mathcal{G}=0.4G$, and $\delta=0.25$. The probability amplitudes of the phonon excitation on sublattices A and B correspond to light and dark gray, respectively. (e) A SSH chain with periodic boundary conditions. \textcolor[rgb]{0,0,0}{A skyrmion qubit is coupled to a site of sublattice A}. (f) \textcolor[rgb]{0,0,0}{The skyrmion coupled to a site of sublattice A} is replaced by a vacancy to obtain a bare \textcolor[rgb]{0,0,0}{phononic open SSH chain with $(2N-1)$ sites}~\cite{1997ShinP249254,2014SirkerP1003210032}.}
    \label{Fig6}
\end{figure*}

We begin to investigate the characteristics of the oscillator array, defining its Hamiltonian as
\begin{equation}
    \hat{H}_{\rm{Ph}}=\sum_n \Omega_m \hat{b}_n^{\dagger}\hat{b}_n+\sum_n G_n\left(\hat{b}_n^{\dagger}\hat{b}_{n+1}+H.c.\right).
    \label{HPh}
\end{equation}
In this case, we consider a spatial period of $2$, which allows us to obtain the Su-Schrieffer-Heeger (SSH) model, and the Hamiltonian of the phonon modes simplifies to
\begin{equation}
    \begin{split}
        \hat{H}_{\rm{Ph}}&=\sum_n \Omega_m \left(\hat{A}_n^{\dagger}\hat{A}_n+\hat{B}_n^{\dagger}\hat{B}_n\right)\\
        &+\sum_n \left(G_1\hat{A}_n^{\dagger}\hat{B}_n+G_2\hat{B}_n^{\dagger}\hat{A}_{n+1}+H.c.\right),
    \end{split}
    \label{HPhSSH}
\end{equation}
where $\hat{A}_n$ ($\hat{B}_n$) is the annihilation operator for phonons in sublattice A (B) [Fig.~\ref{Fig5}(c)]. By altering the voltage $U$, the coupling strengths $G_1\equiv G(1+\delta)$ and $G_2\equiv G(1-\delta)$ can be changed (see Appendix~\ref{sec:appendixF}). Through the Fourier transform
\begin{equation}
    \hat{A}_k=\frac{1}{\sqrt{N}}\sum_{n=1}^{N} e^{-ikn}\hat{A}_n,~\hat{B}_k=\frac{1}{\sqrt{N}}\sum_{n=1}^{N} e^{-ikn}\hat{B}_n,
    \label{FT}
\end{equation}
the Hamiltonian of the phononic bath can be reduced to
\begin{equation}
    \hat{H}_{\rm{ph}}=
    \begin{pmatrix}
        \hat{A}_k^{\dagger} & \hat{B}_k^{\dagger}
    \end{pmatrix}
    \begin{pmatrix}
        \Omega_m & G_1+G_2e^{-ik} \\
        G_1+G_2e^{ik} & \Omega_m
    \end{pmatrix}
    \begin{pmatrix}
        \hat{A}_k \\
        \hat{B}_k
    \end{pmatrix},
    \label{HPhDiag}
\end{equation}
where the periodicity condition is applied. The dispersion relations of the phononic bath are determined by diagonalizing Hamiltonian~(\ref{HPhDiag}), i.e.
\begin{equation}
    \Omega_{\pm}=\pm G\sqrt{2\left(1+\delta^2\right)+2\left(1-\delta^2\right)\cos k}
    \label{DisRela}
\end{equation}
with $\Omega_m$ as the energy reference. These two dispersion relations can be further described as $\Omega_+(k)=-\Omega_-(k)=\Omega(k)$ because they are symmetric with regard to the zero energy. Figure~\ref{Fig6}(a) depicts the dispersion relation with a band gap $4G\vert\delta\vert$. As the squeezed parameter $r$ grows, the bandgap of the phononic bath becomes much larger [Fig.~\ref{Fig6}(b)].
\subsection{\label{sec:IVB}A single skyrmion qubit coupled to the phononic bath}
We  investigate the properties of the hybrid system composed of a single skyrmion qubit and the phononic bath with a topological band structure. The interaction Hamiltonian of the system in $k$-space is
\begin{equation}
    \begin{split}
        \hat{H}_{\rm{int}}&=\frac{\mathcal{G}}{\sqrt{2N}}\sum_{k,n}\hat{\sigma}_+^{n,A}e^{ikx_n}\left(\hat{\alpha}_k-\hat{\beta}_k\right)\\ &+\frac{\mathcal{G}}{\sqrt{2N}}\sum_{k,n}\hat{\sigma}_+^{n,B}e^{i[kx_n-\phi(k)]}\left(\hat{\alpha}_k+\hat{\beta}_k\right)+H.c.,
    \end{split}
    \label{HintkSpace}
\end{equation}
where $\hat{\alpha}_k/\hat{\beta}_k=\left[\pm\hat{A}_k+e^{i\phi(k)}\hat{B}_k\right]/\sqrt{2}$ are the eigenoperators of the phonon modes on a diagonal basis, respectively. The interaction between the skyrmion qubit and the sublattice A (sublattice B) is denoted by $\hat{\sigma}_+^{n,A}$ ($\hat{\sigma}_+^{n,B}$). $x_n$ denotes the position of the cell and $\phi(k)=\arg [G_1+G_2\exp (-ik)]$. When the skyrmion qubit and phononic bath are coupled, the topological features of the phononic bath are inherited via the skyrmion-phonon interaction. When the frequency of the skyrmion qubit equals the zero-energy mode, it becomes an effective boundary of the phonon lattice, confining the phonon excitations to its one side only~\cite{2019BelloP297297,2021KimP1101511015}. The wave function in the single excitation subspace can be expressed as
\begin{equation}
    \vert \psi \rangle=\left(C_e \hat{\sigma}_++\sum_k\sum_{s=A,B}C_{k,s}^m\hat{s}_k^{\dagger}\right)\vert g\rangle\vert {\rm{vac}}\rangle,
\end{equation}
where $C_e$ is the probability amplitude of  the spin in the excited state, and $C_{k,s}^m$ indicates the probability amplitude that the phonon excitation exists in different sublattices when the spin is coupled to the sublattice $m=A/B$ in the $j=0$ cell. The notations $\hat{s}_k^{\dagger}=\{\hat{A}_k^\dagger,\hat{B}_k^\dagger\}$, $\vert g\rangle$, and $\vert {\rm vac}\rangle$ denote the creation operator for phonon modes in $k$ space, the ground state of the skyrmion qubit, and the vacuum state of the phonon mode, respectively. By solving the equation $\hat{H}_{\rm{MTMS}}^{\rm{sq}}\vert \psi_{\rm{BS}}\rangle=E_{\rm{BS}}\vert \psi_{\rm{BS}}\rangle$, we have
\begin{subequations}
    \begin{equation}
        C_{k,A}^A=\frac{\mathcal{G}C_eE_{\rm{BS}}}{2\pi}\int_{-\pi}^{\pi}dk \frac{e^{ikj}}{E_{\rm{BS}}^2-\Omega^2(k)},
        \label{CKAa}
    \end{equation}
    \begin{equation}
        C_{k,A}^B=\frac{\mathcal{G}C_e}{2\pi}\int_{-\pi}^{\pi}dk \frac{\Omega(k)e^{i[kj-\phi(k)]}}{E_{\rm{BS}}^2-\Omega^2(k)},
        \label{CKBa}
    \end{equation}
    \begin{equation}
        C_{k,B}^A=\frac{\mathcal{G}C_e}{2\pi}\int_{-\pi}^{\pi}dk \frac{\Omega(k)e^{i[kj-\phi(k)]}}{E_{\rm{BS}}^2-\Omega^2(k)},
        \label{CKAb}
    \end{equation}
    \begin{equation}
        C_{k,B}^B=\frac{\mathcal{G}C_eE_{\rm{BS}}}{2\pi}\int_{-\pi}^{\pi}dk \frac{e^{ikj}}{E_{\rm{BS}}^2-\Omega^2(k)},
        \label{CKBb}
    \end{equation}
\end{subequations}
where we have employed the Fourier transform to derive the distribution in real space. Figures~\ref{Fig6}(c, d) show the phonon mode localization on the right side of the skyrmion qubit when it is coupled to sublattice A, and on the left side of the skyrmion qubit when it is coupled to sublattice B. Figure~\ref{Fig6}(e) illustrates that the SSH chain can be represented as a circle under periodic boundary conditions. A skyrmion qubit is coupled to \textcolor[rgb]{0,0,1}{one of the sites of sublattice A}. Figure~\ref{Fig6}(f), on the other hand, \textcolor[rgb]{0,0,1}{shows the site of sublattice A coupled to} a skyrmion qubit is replaced with a vacancy. In this case, the skyrmion-phonon bound state obtained in Fig.~\ref{Fig6}(e) is one-to-one mapping with the edge state obtained in Fig.~\ref{Fig6}(f), as demonstrated in Ref.~\cite{2021LeonforteP6360163601}. In other words, the skyrmion-phonon bound state corresponds to the left and right edge state appearing in the topological phase, and the skyrmion coupling induces an effective boundary in the phonon SSH chain~\cite{2021LeonforteP6360163601,2019BelloP297297,2021KimP1101511015,2022DongP2307723077}. It is worth noting that for $\delta=0$, the chain of mechanical oscillators is reduced to a trivial phononic chain, at which case the chirality of the skyrmion-phonon bound state will vanish~\cite{2015DouglasP326331,2023RenP3371733717}.

\begin{figure}
    \centering
    \includegraphics[width=0.48\textwidth]{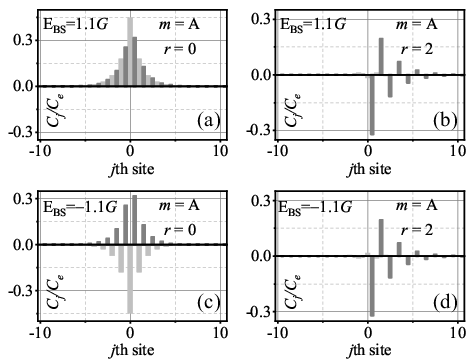}
    \caption{Figures (a, c) show that the chirality of the skyrmion-phonon bound state disappears when $E_{\rm{BS}}\neq 0$ ($r=0$), but it reappears when the squeezed parameter $r$ is adjusted, i.e., the squeezed parameter $r$ can modulate the chirality of the skyrmion-phonon bound state [(b) and (d), $r=2$]. The parameters are $\mathcal{G}=0.4G$ and $\delta=0.25$.}
    \label{Fig7}
\end{figure}

In the preceding discussion, we assumed $E_{\rm{BS}}=0$. Next, we analyze the case $E_{\rm{BS}}\neq0$. Figures~\ref{Fig7}(a, c) illustrate the distribution of phonon excitations in real space for $E_{\rm{BS}}=1.1G$ and $E_{\rm{BS}}=-1.1G$, respectively. It is obvious that when $E_{\rm{BS}}\neq 0$, the chirality of the skyrmion-phonon bound state disappears. When we increase the squeezed parameter $r$, the chirality of the skyrmion-phonon bound state returns, as can be seen in Figs.~\ref{Fig7}(b, d). In other words, by modulating the two-phonon drive to the mechanical oscillator, we can manipulate the system chirality. The phenomenon can be explained by Fig.~\ref{Fig6}(b). When the squeezed parameter $r=0$, the nonzero $E_{\rm{BS}}$ cannot satisfy the condition $E_{\rm{BS}}\approx 0$, and then the chirality will disappear, but as the squeezed parameter $r$ increases, the bandgap also increases, and then the condition $E_{\rm{BS}}\approx 0$ approximates to be valid and the chirality reappears. For the case $m=B$, the local direction of the skyrmion-phonon bound state will be reversed, as shown in Fig.~\ref{Fig6}(c, d).
\subsection{\label{sec:IVC}Multiple skyrmion qubits coupled to the topological phonon bath}
Subsequently, we investigate effective skyrmion-skyrmion interactions that are mediated by chiral skyrmion-phonon bound states. The interaction between skyrmion qubits is generated by exchanging virtual phonons. For $\delta > 0$ and with Markov approximation, the Hamiltonian of the effective skyrmion-skyrmion interaction can be expressed as (see Appendix~\ref{sec:appendixG})~\cite{2022GongP5351753517,2022DongP2307723077,2019BelloP297297}
\begin{equation}
    \hat{H}_{\rm{SS}}=-\sum_{i<j}G_{i,j}\left(\hat{\sigma}_+^i\hat{\sigma}_-^j+H.c.\right),
    \label{HSS}
\end{equation}
where the coupling strength is defined as $G_{i,j}^{AA/BB}=0$ and
\begin{equation}
    G_{i,j}^{AB}=\left\{
    \begin{split}
        &\frac{\mathcal{G}^2\left(-1\right)^{x_{ij}}}{G\left(1+\delta\right)}\left(\frac{1-\delta}{1+\delta}\right)^{x_{ij}}, x_{ij}\geq0, \\
        &0, x_{ij} < 0.
    \end{split}
    \right.
    \label{GijA}
\end{equation}
\begin{figure}
    \centering
    \includegraphics[width=0.48\textwidth]{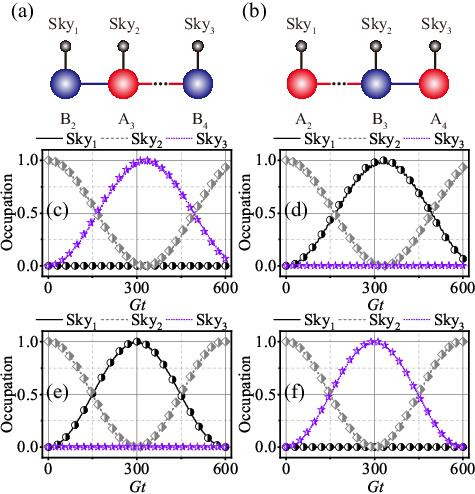}
    \caption{(a) and (b) depict two models with skyrmion qubits coupled to distinct sublattices. The dynamical evolution of the corresponding models is shown in (c) and (d), and it can be seen that altering the coupling points can modulate the skyrmion-skyrmion interaction. In (c) and (d), we assume that $\mathcal{G}=0.1G$ and $\delta=0.25$. (e) and (f) correspond to (c) and (d), respectively, and by adjusting the parameter $\delta=0.25$ to $\delta=-0.9$, the chiral skyrmion-skyrmion interactions are modified, i.e., the skyrmion-skyrmion interactions are controllable by modifying the phonon hopping rate. The scattered dots in (c, d, e, f) illustrate the results of the dynamics simulations obtained using the analytical solution~(\ref{GijA}), which are consistent with the simulations using the original Hamiltonian~(\ref{HMTMSSqr}).}
    \label{Fig8}
\end{figure}

Then we perform numerical simulations for the dynamics of the hybrid system. Three skyrmion qubits are separately coupled to three mechanical oscillators in a mechanical oscillator array consisting of 10 cantilevers [Figs.~\ref{Fig8}(a, b)]. The effective skyrmion-skyrmion interaction in the hybrid quantum system is chiral because it is generated by a chiral skyrmion-phonon bound state. It can be adjusted by altering the position of the skyrmion-coupled sublattice and the hopping rate between phonons. The initial state of the three qubits is $\vert\psi_{\rm{Sky}_1}\rangle\vert\psi_{\rm{Sky}_2}\rangle\vert\psi_{\rm{Sky}_3}\rangle=\vert g\rangle\vert e\rangle\vert g\rangle$ and that of the phonon is $\vert{\rm{vac}}\rangle$. First, we analyze how to modulate the effective skyrmion-skyrmion coupling by adjusting the coupling position of the skyrmion qubits. As indicated in Fig.~\ref{Fig8}(a), skyrmions 1 (\ce{Sky_1}), 2 (\ce{Sky_2}), and 3 (\ce{Sky_3}) are coupled to sublattices $\ce{B_2}$, $\ce{A_3}$, and $\ce{B_4}$, respectively; however, in the second scenario, \ce{Sky_1}, \ce{Sky_2}, and \ce{Sky_3} are coupled to sublattices $\ce{A_2}$, $\ce{B_3}$, and $\ce{A_4}$, respectively [Fig.~\ref{Fig8}(b)]. Figures~\ref{Fig8}(c) [(e)] and (d) [(f)] illustrate the simulation results for the two cases, respectively. Because of the chirality of the skyrmion-phonon bound state, there is only an interaction between \ce{Sky_2} (\ce{Sky_1}) and \ce{Sky_3} (\ce{Sky_2}), without coupling to \ce{Sky_1} (\ce{Sky_3}), as shown in Fig.~\ref{Fig8}(c) [(e)]. When the position of the coupling of the skyrmions is changed, however, there is an interaction between \ce{Sky_1} (\ce{Sky_2}) and \ce{Sky_2} (\ce{Sky_3}) without coupling to \ce{Sky_3} (\ce{Sky_1}), as illustrated in Fig.~\ref{Fig8}(d) [(f)]. The second approach modulates the effective skyrmion-skyrmion interactions by modifying the hopping rate between phonons, which can be accomplished by changing the voltage $U$. In Fig.~\ref{Fig8}(c) [(d)], we take $\delta=0.25$, and then $G_1>G_2$, indicating that there is an interaction only between \ce{Sky_2} (\ce{Sky_1}) and \ce{Sky_3} (\ce{Sky_2}), but when $\delta=-0.9$, we have $G_1<G_2$, implying that there is an interaction only between \ce{Sky_1} (\ce{Sky_2}) and \ce{Sky_2} (\ce{Sky_3}), as illustrated in Fig.~\ref{Fig8}(e) [(f)]. In other words, by merely altering the voltage $U$, we can modulate the chirality of the skyrmion-skyrmion interaction.

\section{\label{sec:VI}Experimental feasibility}
In the following, the experimental feasibility of the suggested hybrid quantum system is analyzed. Both experimental and theoretical investigations of skyrmions have made significant progress. Skyrmions are predicted to exist in triangular-lattice frustrated magnets \ce{NiGa_2S_4}~\cite{2005NakatsujiP16971700,2010StockP3740237402}, \ce{Fe_xNi_{1-x}Br_2}~\cite{1976DayP24812481,1982RegnaultP12831290,1985MooreP2341}, \ce{\alpha-NaFeO_2}~\cite{2007McQueenP2442024420,2014TeradaP184421184421}, and square-lattice frustrated magnets \ce{Pb_2VO(PO_4)_2}~\cite{2021UkpongP17701770}. Bloch skyrmions have been discovered experimentally in triangular-lattice \ce{Gd_2PdSi_3}~\cite{2019KurumajiP914918} and \ce{Gd_3Ru_4Al_2}~\cite{2019HirschbergerP58315831} via the topological Hall effect. Here, we suppose that the parameters of the skyrmion are the effective spin $\bar{S}=20$, the lattice spacing $a=0.5~\rm{nm}$, the interaction strength $\mathcal{J}_1=1~\rm{meV}$, the anisotropy energy $0.15~\rm{meV}$, the applied external magnetic field $35~\rm{mT}$, and the electric polarization $P_E=0.2~\rm{C/m}$~\cite{2021PsaroudakiP6720167201,2022PsaroudakiP104422104422}. To manipulate the resonance frequency of the skyrmion qubit, the electric field gradient $80~\rm{V/m}$ is added. Based on the parameters mentioned here, we can calculate that the radius of the skyrmion is $\sim 3~\rm{nm}$~\cite{2023XiaP106701106701} and the resonance frequency of the skyrmion qubit is $\omega_q/2\pi\approx2~\rm{GHz}$.

Here, we assume that the geometric parameter of the cantilever is $(l,w,t)=(5.6~\rm{\mu m}, 0.05~\rm{\mu m}, 0.04~\rm{\mu m})$. With the cantilever mass density $\rho=2329~\rm{kg/m^3}$ and the Young's modulus $E=1.3\times 10^{11}~\rm{Pa}$~\cite{2019DongP4382543825,2020LiP153602153602}, we derive the phonon resonance frequency $\omega_m/2\pi=3.516\times t \sqrt{E/12\rho}/l^2\approx 10~\rm{MHz}$. The magnetic tip is assumed to be composed of a uniformly magnetized truncated cone and a non-magnetized protective layer~\cite{2009DegenP13131317}, with the upper bottom surface radius, lower bottom surface radius, magnetized layer thickness, and non-magnetized layer thickness being $r_a=40~\rm{nm}$, $r_b=160~\rm{nm}$, $h_{\rm{Tip}}=180~\rm{nm}$, and $s_{\rm{NM}}=10~\rm{nm}$, respectively. At a distance of $20~\rm{nm}$ from the magnetic tip, the magnetic field gradient is $G\approx1.74\times 10^7~\rm{T/m}$, with the tip magnetization strength $\mu_0 M_s=2.4~\rm{T}$~\cite{2009DegenP13131317}. We can calculate the skyrmion-phonon coupling strength $\lambda_{\rm{TS}}/2\pi\approx 3.56~\rm{MHz}$. In this case, we suppose that the original phonon dissipation is $1~\rm{kHz}$. Taking into account the parametric amplification of dissipation, we eventually assume that the phonon dissipation is $\gamma_m/2\pi=0.1~\rm{MHz}$. The dissipation of the skyrmion qubit is $\gamma_{\rm{Sky}}/2\pi=1~\rm{MHz}$. The cooperativity of the system is $\mathcal{C}_{\rm{TS}}=4\lambda_{\rm{TS}}^2/(\gamma_m\gamma_{\rm{Sky}})\approx 507 \gg 1$, i.e., the system can reach the strong coupling region~\cite{2020ClerkP257267}.

\section{\label{sec:VII}Conclusion}
We propose a hybrid quantum system composed of skyrmion qubits and nanomechanical cantilevers, and show that the coherent coupling between them can reach the strong coupling regime. It is feasible to modulate the skyrmion qubit by employing phonons as quantum interfaces. Firstly, we show the enhancement and modulation of the skyrmion-phonon coupling strength by modulating the two phonon drive. Even ultrastrong coupling of the system can be obtained by increasing the squeezed parameter $r$. Furthermore, by adiabatically eliminating the phonon modes, we establish strong skyrmion-skyrmion interactions. When considering a coupled resonator array, the long-range skyrmion-skyrmion interactions by exchanging virtual phonons can be realized~\cite{2023RenP3371733717}. More interestingly,  when a skyrmion qubit is coupled to a topological oscillator array, we can obtain a chiral skyrmion-phonon bound state, and we can alter the chirality by changing the squeezed parameter $r$. Utilizing the chiral skyrmion-phonon bound state, chiral skyrmion-skyrmion interactions can be obtained. It can be modulated by altering the coupling position of the skyrmion qubit or the hopping rate between phonons. The highly controllable hybrid quantum system suggested here offers a feasible platform for quantum technologies and quantum information.

\begin{acknowledgments}
This work is supported by the National Natural Science Foundation of China under Grants No.
12375018 and No. 92065105.
\end{acknowledgments}

\appendix
\section{\label{sec:appendixA}The classical skyrmion}
A skyrmion is a centrosymmetric vortex-structured spin texture. The skyrmion in the frustrated material, in particular, possesses an internal degree of freedom: helicity $\varphi_0$. According to the Ginzburg-Landau theory~\cite{2016LinP6443064430}, the Hamiltonian of the skyrmion in frustrated material can be represented as
\begin{equation}
    \mathcal{H}=\int d \widetilde{\boldsymbol{r}} \left[-\frac{\mathcal{J}_1^2}{2}\left(\nabla_{\widetilde{\boldsymbol{r}}}\boldsymbol{s}\right)^2+\frac{\mathcal{J}_2a^2}{2}\left(\nabla^2_{\widetilde{\boldsymbol{r}}}\boldsymbol{s}\right)^2-\frac{H}{a^2}s_z+\frac{K}{a^2}s_z^2\right].
\end{equation}
The vector $\boldsymbol{s}$ is the normalized spin vector, defined as $\boldsymbol{s}=\boldsymbol{S}/\vert \boldsymbol{S}\vert=(s_x,s_y,s_z)$, employed to describe the spin direction at location $\widetilde{\boldsymbol{r}}=(\widetilde{\rho},\phi)$, where $\widetilde{\rho}$ and $\phi$ are the polar coordinate variables. We have assumed that the skyrmion is in the $xy$ plane, with its center at the coordinate origin. For a classical skyrmion, $\boldsymbol{s}=[\sin\Theta(\widetilde{\rho})\cos\Phi, \sin\Theta(\widetilde{\rho})\sin\Phi,\cos\Theta(\widetilde{\rho})]$ with $\Phi=\phi+\varphi_0$. Minimizing the energy $\mathcal{H}$ yields the classical stationary solutions $\Theta_0$ and $\Phi_0$. Here we take the approximate solution $\Theta_0=\pi/\sqrt{\rho^2+1}\exp(-\mathcal{Y}_{\rm{Re}}\rho)\cos(-\mathcal{Y}_{\rm{Im}}\rho)$ with $\mathcal{Y}_{\rm{Re}}=\rm{Re}(\mathcal{Y})$, $\mathcal{Y}_{\rm{Im}}=\rm{Im}(\mathcal{Y})$, $\mathcal{Y}=\sqrt{-1+\widetilde{\mathcal{Y}}}/\sqrt{2}$, $\widetilde{\mathcal{Y}}=\sqrt{1-4(h+\kappa_z)}$, $h=H/\mathcal{J}_1$, and $\kappa_z=K/\mathcal{J}_1$.

\section{\label{sec:appendixB}The skyrmion qubit}
The collective coordinate quantization approach has been thoroughly discussed in Ref.~\cite{1975GoldstoneP14861498,1994DoreyP35983611,2021PsaroudakiP6720167201,2022PsaroudakiP104422104422}, and the main process of the method will be described here. The partition function of the skyrmion studied here is
\begin{equation}
    \mathcal{Z}=\int \mathcal{D}\boldsymbol{s}\exp\left[i\mathcal{S}\left(\boldsymbol{s},\dot{\boldsymbol{s}}\right)\right]
\end{equation}
with action $\mathcal{S}=\int dt \mathcal{L}$, where $\mathcal{L}$ denotes Lagrangian
\begin{equation}
    \mathcal{L}=\bar{S}\int d\boldsymbol{r} \left[\mathfrak{A}\left(\boldsymbol{s}\right)\cdot \dot{\boldsymbol{s}}-\mathcal{F}\right]
    \label{L}
\end{equation}
with gauge potential $\mathfrak{A}=\left\{[1-\widetilde{e}_{\Phi}\cdot (e_{\Phi}\times\boldsymbol{s})]/(\widetilde{e}_{\Phi})\cdot \boldsymbol{s}\right\}e_{\Phi}$. $\mathfrak{A}(\boldsymbol{s})\cdot\boldsymbol{s}=(1-\cos\Theta)\dot{\Phi}$ can be obtained by using $\widetilde{e}_\Phi=(\cos\Phi,\sin\Phi,0)$ and $e_{\Phi}=(-\sin\Phi,\cos\Phi,0)$. $\mathcal{F}$ is the energy density functional, which is defined as
\begin{equation}
    \mathcal{F}=-\frac{\mathcal{J}_1^2}{2}\left(\nabla_{\widetilde{\boldsymbol{r}}}\boldsymbol{s}\right)^2+\frac{\mathcal{J}_2a^2}{2}\left(\nabla^2_{\widetilde{\boldsymbol{r}}}\boldsymbol{s}\right)^2-\frac{H}{a^2}s_z+\frac{K}{a^2}s_z^2.
\end{equation}
Based on the global symmetry of the model, we can get $\boldsymbol{s}\rightarrow\mathcal{M}(\varphi_0)\boldsymbol{s}$, where $\mathcal{M}(\varphi_0)$ is given by
\begin{equation}
    \mathcal{M}(\varphi_0)=
    \begin{pmatrix}
        \cos\varphi_0 & -\sin\varphi_0 & 0 \\
        \sin\varphi_0 & \cos\varphi_0 & 0 \\
        0 & 0 & 1
    \end{pmatrix}.
\end{equation}
By introducing the vector $\boldsymbol{n}=\sqrt{1-\cos\Theta}\boldsymbol{s}/\sin\Theta$, the gauge potential can be redefined as $\boldsymbol{\mathfrak{A}}(\boldsymbol{n})=\partial_{\Phi}\boldsymbol{n}$. Based on the collective coordinate quantization method, we introduce $\delta$ constraints~\cite{1994DoreyP35983611,1996BraunP32373255}
\begin{equation}
    \int \mathcal{D}\varphi_0\mathcal{D}\mathfrak{S}_z J_{\varphi_0}J_{\mathfrak{S}_z}\delta(F_1)\delta(F_2)=1
\end{equation}
with
\begin{equation}
    \begin{split}
        F_1&=\int d\boldsymbol{r} \boldsymbol{\mathfrak{A}}(\boldsymbol{n}_0)\cdot\left(\widetilde{\boldsymbol{n}}-\boldsymbol{n}_0\right), \\
        F_2&=\frac{1}{\Lambda}\int d\boldsymbol{r} \boldsymbol{\mathfrak{A}}(\boldsymbol{n}_0)\cdot\left[\widetilde{\boldsymbol{\mathfrak{A}}}(\boldsymbol{n})-\boldsymbol{\mathfrak{A}}(\boldsymbol{n}_0)\right], \\
        J_{\varphi_0}&=\frac{\delta F_1}{\delta\varphi_0},~
        J_{\mathfrak{S}_z}=\frac{\delta F_2}{\delta\mathfrak{S}_z},
    \end{split}
\end{equation}
which eliminate zero modes. The Jacobians of the transformation are $J_{\varphi_0}$ and $J_{\mathfrak{S}_z}$. We express the field as its classical solution plus its corresponding quantum fluctuations, introducing $\boldsymbol{n}=\widetilde{\boldsymbol{n}}_0+\widetilde{\boldsymbol{\gamma}}$ and $\boldsymbol{\mathfrak{A}}(\boldsymbol{n})=c\widetilde{\boldsymbol{\mathfrak{A}}}(\boldsymbol{n}_0)+\widetilde{\boldsymbol{\zeta}}$. The parameter $c$ ensures that the transformations described above are canonical. And applying the constraint $P-\mathfrak{S}_z=F_2$, we can get
\begin{equation}
    c=\frac{\mathfrak{S}_z-\int d\boldsymbol{r} \widetilde{\boldsymbol{\zeta}}\cdot\partial_{\phi}\boldsymbol{n}}{\int d\boldsymbol{r} \widetilde{\boldsymbol{\mathfrak{A}}}(\boldsymbol{n}_0)\cdot\partial_{\phi}\boldsymbol{n}}.
\end{equation}
In other words, we can derive the path integral in phase space retaining the canonical form $\int dt d\boldsymbol{r}\boldsymbol{\mathfrak{A}}(\boldsymbol{n})\cdot\dot{\boldsymbol{n}}=\int dt \mathfrak{S}_z\dot{\varphi}_0+\int dt d\boldsymbol{r} \boldsymbol{\zeta}\cdot\dot{\boldsymbol{\gamma}}$. In order to simplify the analysis of the transformation of the energy functional $\mathcal{F}$, the transformation
\begin{equation}
    \begin{split}
        \widetilde{\Pi}(\boldsymbol{r},t)&=\frac{\mathfrak{S}_z-\int d\boldsymbol{r}\eta(\boldsymbol{r,t})\partial_{\phi}\Phi(\boldsymbol{r},t)}{\Lambda}\widetilde{\Pi}_0(\boldsymbol{r},t)+\eta(\boldsymbol{r},t), \\
        \Phi(\boldsymbol{r},t)&=\Phi_0\left[\boldsymbol{r},\varphi_0(t)\right]+\xi(\boldsymbol{r},t)
    \end{split}
    \label{CCQ}
\end{equation}
is introduced, ensuring that the canonical form of the Wess-Zumino (WZ) term $\int d\boldsymbol{r} \widetilde{\Pi}\dot{\Phi}=\mathfrak{S}_z\dot{\varphi}_0+\int d\boldsymbol{r} \eta\dot{\xi}$ is maintained. Based on $\mathcal{S}=\int dt \mathcal{L}$ and Eq.~(\ref{L}), we can derive the real-time action as~\cite{1996BraunP32373255}
\begin{equation}
    \mathcal{S}=\frac{\bar{S}}{a^2}\int d\widetilde{t}d\widetilde{\boldsymbol{r}} \dot{\Phi}\left(1-\Pi\right)-\int d\widetilde{t}\mathcal{H}_{\mathfrak{S}_z},
    \label{Action}
\end{equation}
where $\dot{\Phi}\equiv\partial_t\Phi$ and $\Pi\equiv \cos\Theta$. Please keep in mind that we only discuss the construction of $\mathfrak{S}_z$ qubits, which can be described by the Hamiltonian
\begin{equation}
    \mathcal{H}_{\mathfrak{S}_z}=\bar{S}\int d\widetilde{\boldsymbol{r}}\left[-\frac{H}{a^2}s_z+\frac{K}{a^2}s_z^2-EP_Ea\hat{\boldsymbol{e}}_z\cdot\boldsymbol{P}\right],
    \label{EFAll}
\end{equation}
where the electric polarization $\boldsymbol{P}=\hat{\boldsymbol{e}}_x\times(\boldsymbol{s}\times\partial_{\widetilde{x}}\boldsymbol{s})+\hat{\boldsymbol{e}}_y\times(\boldsymbol{s}\times\partial_{\widetilde{y}}\boldsymbol{s})$. To manipulate the energy levels of qubits, we introduce an external electric field in the $z$-direction. Defining dimensionless parameters $\mathcal{J}_{\Lambda}=\mathcal{J}_1$, $\ell=\sqrt{\mathcal{J}_2/\mathcal{J}_1}$, $\widetilde{\boldsymbol{r}}=\boldsymbol{r}\ell a$, and $\widetilde{t}=t/\mathcal{J}_{\Lambda}$, one can obtain
\begin{equation}
    \mathcal{S}=\bar{S}\int dtd\boldsymbol{r}\dot{\Phi}\left(1-\Pi\right)-\int dt \mathcal{H}_{\mathfrak{S}_z},
    \label{ActionDimen}
\end{equation}
where
\begin{equation}
    \mathcal{H}_{\mathfrak{S}_z}=\int d\boldsymbol{r}\left(-\bar{h}_z s_z+\bar{\kappa}_z s_z^2-\bar{\varepsilon}_z\hat{\boldsymbol{e}}_z\cdot\bar{\boldsymbol{P}}\right).
    \label{HSzDimless}
\end{equation}
The coefficients in Eq.~(\ref{HSzDimless}) are defined as $\bar{h}_z=H\bar{S}/\mathcal{J}_{\Lambda}$, $\bar{\kappa}_z=K\bar{S}/\mathcal{J}_{\Lambda}$, $\bar{\varepsilon}_z=a^3E P_E\bar{S}/\mathcal{J}_{\Lambda}$, and $\bar{\boldsymbol{P}}=\hat{\boldsymbol{e}}_x\times(\boldsymbol{s}\times\partial_x\boldsymbol{s})+\hat{\boldsymbol{e}}_y\times(\boldsymbol{s}\times\partial_y\boldsymbol{s})$~\cite{2021PsaroudakiP6720167201,2018TakeiP6440164401}. According to transformation~(\ref{CCQ}), the Hamiltonian~(\ref{HSzDimless}) can be expressed as
\begin{equation}
    \mathcal{H}_{\mathfrak{S}_z}=\bar{\bar{\kappa}}_z\mathfrak{S}_z^2-\bar{\bar{h}}_z\mathfrak{S}_z-\bar{\bar{\varepsilon}}_z\cos\varphi_0,
    \label{HSz}
\end{equation}
where $\varphi_0$ and $\mathfrak{S}_z$ denote the collective coordinate and its conjugate momentum, respectively. The terms $\eta$ and $\xi$, which are associated with magnon fluctuations, have been ignored here. The coefficients in Hamiltonian~(\ref{HSz}) are defined as
\begin{equation}
    \begin{split}
        \bar{\bar{\kappa}}_z&=\bar{\kappa}_z\int d\boldsymbol{r}\frac{\left(1-\cos\Theta_0\right)^2}{\int d\boldsymbol{r}\left(1-\cos\Theta_0\right)^2}, \\
        \bar{\bar{h}}_z&=\bar{h}_z, \\
        \bar{\bar{\varepsilon}}_z&=\bar{\varepsilon}_z\int d\boldsymbol{r}\left[\partial_{\rho}\Theta_0+\frac{\sin(2\Theta_0)}{2\rho}\right].
    \end{split}
\end{equation}
The operatorization of the collective coordinate and the corresponding momentum gives $\hat{\varphi}_0$ and $\hat{\mathfrak{S}}_z$, satisfying $\hat{\mathfrak{S}}_z\vert s\rangle=s\vert s\rangle$, $\hat{\varphi}_0\vert \varphi_0\rangle=\varphi_0\vert \varphi_0\rangle$, $e^{\pm i\hat{\varphi}_0}\vert s\rangle=\vert s\pm 1\rangle$, and $\left[\hat{\varphi}_0,\hat{\mathfrak{S}}_z\right]=i$. The Hamiltonian of the skyrmion qubit can thus be reduced to
\begin{equation}
    \hat{\mathcal{H}}_{\mathfrak{S}_z}=\bar{\bar{\kappa}}_z\hat{\mathfrak{S}}_z^2-\bar{\bar{h}}_z\hat{\mathfrak{S}}_z-\bar{\bar{\varepsilon}}_z\cos\hat{\varphi}_0.
    \label{HSzQuantum}
\end{equation}
The eigenstates and eigenenergies of the skyrmion qubit can be derived by solving the Schr\"odinger equation $\hat{\mathcal{H}}_{\mathfrak{S}_z}\Psi_s(\varphi_0)=\mathfrak{E}_s\Psi_s(\varphi_0)$. Substituting the Hamiltonian~(\ref{HSzQuantum}) into the Schr\"odinger equation, the damped Mathieu equation can be obtained~\cite{2021NwaigweP6270262702,2021ChakiP600605}
\begin{equation}
    \left[\partial_{\varphi_0}^2-i\widetilde{h}\partial_{\varphi_0}+\frac{\mathfrak{E}_s}{\bar{\bar{\kappa}}_z}+\frac{\bar{\bar{\varepsilon}}_z}{\bar{\bar{\kappa}}_z}\cos\varphi_0\right]\Psi_s(\varphi_0)=0,
    \label{DME}
\end{equation}
where $\Psi_s(\varphi_0)=\langle\varphi_0\vert s\rangle$ and $\Psi_s(\varphi_0)=\Psi_s(\varphi_0+2\pi)$. Utilizing the Liouville transformation $\Psi_s(\varphi_0)=\psi_s(\varphi_0)\exp(i\widetilde{h}\varphi_0/2)$~\cite{2021NwaigweP6270262702}, the damped Matthieu equation Eq.~(\ref{DME}) can be simplified to the standard Matthieu equation
\begin{equation}
    \partial_{\varphi_0}^2\psi_s(\varphi_0)+\left[\alpha_x+q_x\cos\varphi_0\right]\psi_s(\varphi_0)=0
    \label{StandardMEqq}
\end{equation}
with eigenvalues and eigenfunctions
\begin{equation}
    \begin{split}
        \alpha_x &= \frac{\mathfrak{M}_A(\mathfrak{N},-2q_x)}{4}, \\
        \psi_s(\varphi_0)&=\sum_{j=C/S}\mathfrak{C}_j\mathfrak{M}_j\left(4\alpha_x,-2q_x,\varphi_0/2\right),
    \end{split}
\end{equation}
where $\alpha_x=\widetilde{h}^2/4+\mathfrak{E}_s/\bar{\bar{\kappa}}_z$ and $q_x=\bar{\bar{\varepsilon}}_z/\bar{\bar{\kappa}}_z$. Odd solutions, even solutions, and characteristic indices are denoted by $\mathfrak{M}_S$, $\mathfrak{M}_C$, and $\mathfrak{N}$, respectively. Because it is demonstrated in Ref.~\cite{2021PsaroudakiP6720167201} that the energy level of Hamiltonian~(\ref{HSzQuantum}) is inhomogeneous, we can truncate the energy level of the $\mathfrak{S}_z$ qubit to the subspace $\{\vert0\rangle,\vert 1\rangle\}$. The Hamiltonian can then be reduced further as Eq.~(\ref{HSky}).

\section{\label{sec:appendixC}Magnetic field generated by the magnetic tip}
\begin{figure}
    \centering
    \includegraphics[width=0.3\textwidth]{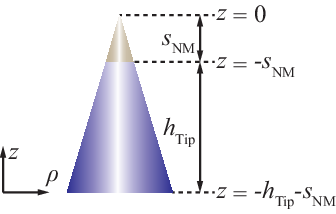}
    \caption{A schematic diagram of the magnetic tip.}
    \label{Fig-Tip}
\end{figure}
The magnetic tip is modeled here as a uniformly magnetized truncated cone with a non-magnetized protecting layer in the top cross-section (Fig.~\ref{Fig-Tip})~\cite{2009DegenP13131317}. The $z$ component of the magnetic field generated by the magnetic tip is
\begin{subequations}
    \begin{equation}
        B_z\left(\rho,z\right)=\frac{\mu_0 M_s}{2\pi}\int\limits_{-h_{\rm{Tip}}-s_{\rm{NM}}}^{-s_{\rm{NM}}}f_{B_z,1}f_{B_z,2}dz^{\prime},
    \end{equation}
    \begin{equation}
        f_{B_z,1}=\frac{1}{\left[\left(R(z^{\prime})+\rho\right)^2+\left(z-z^{\prime}\right)^2\right]^{1/2}},
    \end{equation}
    \begin{equation}
        f_{B_z,2}=\left[\frac{R^2\left(z^{\prime}\right)-\rho^2-\left(z-z^{\prime}\right)^2}{\left(R(z^{\prime})+\rho\right)^2+\left(z-z^{\prime}\right)^2}E\left(q\right)+K\left(q\right)\right],
    \end{equation}
\end{subequations}
and the radius of the magnetic tip at position $z^{\prime}$ is defined as
\begin{equation}
    R\left(z^{\prime}\right)=\frac{r_a-r_b}{h_{\rm{Tip}}}\left(z^{\prime}+s_{\rm{NM}}\right)+r_a.
\end{equation}
The upper and lower surface radiuses of the magnetic tip are represented by the symbols $r_a$ and $r_b$, respectively. The height of the uniform magnetization of the tip and the thickness of the non-magnetized layer are denoted by $h_{\rm{Tip}}$ and $s_{\rm{NM}}$, respectively. The complete elliptic integrals of the first and second kinds are given by
\begin{subequations}
    \begin{equation}
        E\left(q\right)=\int\limits_{0}^{\pi/2}\frac{d\theta}{\sqrt{1-q^2\sin^2\theta}},
    \end{equation}
    \begin{equation}
        K\left(q\right)=\int\limits_{0}^{\pi/2}\sqrt{1-q^2\sin^2\theta}d\theta.
    \end{equation}
\end{subequations}
After some computations, we discover that the gradient of the magnetic field generated by the magnetic tip at $z$ can be approximated as that produced by the magnetic tip at $(0, z)$ in the range we are interested in. The magnetic field produced by the magnetic tip at $z$ can then be simplified to
\begin{subequations}
    \begin{equation}
        B_z\left(0,z\right)=\frac{\mu_0 M_s}{2\pi}\int\limits_{-h_{\rm{Tip}}-s_{\rm{NM}}}^{-s_{\rm{NM}}}f_{B_z,1}^0f_{B_z,2}^0dz^{\prime},
        \label{BZZERO}
    \end{equation}
    \begin{equation}
        f_{B_z,1}^0=\frac{1}{\left[R^2(z^{\prime})+\left(z-z^{\prime}\right)^2\right]^{1/2}},
    \end{equation}
    \begin{equation}
        f_{B_z,2}^0=\frac{\pi}{2}\left[\frac{R^2\left(z^{\prime}\right)-\left(z-z^{\prime}\right)^2}{R^2(z^{\prime})+\left(z-z^{\prime}\right)^2}+1\right].
    \end{equation}
\end{subequations}
Taking the first-order derivative of Eq.~(\ref{BZZERO}) results in the magnetic field gradient
 \begin{subequations}
    \begin{equation}
        \frac{\partial B_z\left(0,z\right)}{\partial z}=\frac{\mu_0 M_s}{4}\int\limits_{-h_{\rm{Tip}}-s_{\rm{NM}}}^{-s_{\rm{NM}}}\left[-\frac{P}{R_z^{1/2}}-\frac{Q}{R_z^{3/2}}\right]dz^{\prime},
    \end{equation}
    \begin{equation}
        \begin{split}
            P&=\frac{2\left[R^2(z^{\prime})+\left(z-z^{\prime}\right)^2\right]\left(z-z^{\prime}\right)}{\left[R^2(z^{\prime})+\left(z-z^{\prime}\right)^2\right]^2}\\ &+\frac{2\left(z-z^{\prime}\right)}{R^2(z^{\prime})+\left(z-z^{\prime}\right)^2},
        \end{split}
    \end{equation}
    \begin{equation}
        Q=\left[1+\frac{R^2(z^{\prime})-\left(z-z^{\prime}\right)^2}{R^2(z^{\prime})+\left(z-z^{\prime}\right)^2}\right]\left(z-z^{\prime}\right),
    \end{equation}
    \begin{equation}
        R_z^{1/2}=\left[R^2(z^{\prime})+\left(z-z^{\prime}\right)^2\right]^{1/2},
    \end{equation}
    \begin{equation}
        R_z^{3/2}=\left[R^2(z^{\prime})+\left(z-z^{\prime}\right)^2\right]^{3/2}.
    \end{equation}
\end{subequations}
Considering the magnetic tip parameters $r_a=40~\rm{nm}$, $r_b=160~\rm{nm}$, $h_{\rm{Tip}}=180~\rm{nm}$, $s_{\rm{NM}}=10~\rm{nm}$, and $\mu_0 M_s=2.4~\rm{T}$, the magnitude of the magnetic field gradient generated by the magnetic tip at a distance of $z=20~\rm{nm}$ from it is $G\approx 1.74\times 10^7~\rm{T/m}$.

\section{\label{sec:appendixD}Nonlinear resources induced by linear drives}
In this section we derive the two-phonon drive induced by the linear drive. The mechanical oscillator driven by a voltage can be described by the Hamiltonian
\begin{equation}
    \hat{H}_m=\frac{\hat{p}^2}{2m}+\frac{1}{2}k(t)\hat{z}^2,
\end{equation}
where $k(t)=k_0+k_E(t)$ and $k_E(t)\equiv\partial F_e/\partial z$. The electrostatic force generated by the electrode is defined as $F_e=\partial (C_r V)^2/(2\partial z)$, where $C_r=\epsilon S/(d+z)=\epsilon_0 \epsilon_r S/(d+z)$ denotes the capacitance of a parallel plate capacitor with the effective area $S$, the vacuum permittivity $\epsilon_0$, the relative permittivity $\epsilon_r$, and the distance between the two plates $(d+z)$. Assuming $V(t)=V_0+V_p\cos(2\omega_E t)$, we can derive the time-dependent stiffness coefficient $k_E(t)=\Delta k_E\cos(2\omega_E t)$ with $\Delta k_E=2\epsilon_0 \epsilon_r S V_0 V_p/d^3$.

After driving the mechanical oscillator, the Hamiltonian of the hybrid system is given by
\begin{equation}
    \begin{split}
        \hat{H}_{\rm{TTS}}&=\frac{\widetilde{\omega}_q}{2}\hat{\sigma}_z+\omega_{m}\hat{a}^{\dagger}\hat{a}+\bar{\lambda}_{\rm{TS}}\left(\hat{a}+\hat{a}^{\dagger}\right)\hat{\sigma}_x \\&-\Omega_E\cos(2\omega_E t)\left(\hat{a}^{\dagger}+\hat{a}\right)^2,
        \label{HTTSD}
    \end{split}
\end{equation}
where $\Omega_E/2\pi=-\Delta k_E z_0^2/(2\pi\hbar)$. Transforming to a rotating frame with the frequency $\omega_E$ and utilizing the rotating wave approximation, one can get Eq.~(\ref{HTTSDD}).

\section{\label{sec:appendixE}The mechanical oscillator array}
In this section we will analyze in detail the interactions between the different oscillators in the array of oscillators. The different resonators in the array can couple to each other by capacitances, as shown in Fig.~\ref{Fig5}(a), and the coupling strength can be calculated by computing the electrostatic energy of the circuit. The interaction between the oscillators is given by the Hamiltonian~\cite{2010RablP602608}
\begin{equation}
    \hat{H}_{\rm{pp}}=g_{ij}\left(\hat{a}_i^{\dagger}+\hat{a}_i\right)\left(\hat{a}_j^{\dagger}+\hat{a}_j\right)
    \label{HPPOrigin}
\end{equation}
with the coupling strength denoted as
\begin{equation}\label{g}
    g_{ij}\equiv g=z_0^2\frac{\partial ^2 W_{\rm{el}}}{\partial z_i \partial z_j}\vert_{z_i=0}.
\end{equation}
$z_0$ is the zero-point fluctuation of the mechanical oscillator, and $W_{\rm{el}}=-U^2/2C_{\Sigma}C_W/(C_{\Sigma}+C_W)$ is the electrostatic energy of the circuit with $C_{\Sigma}=C(1-z_i/h)+C(1-z_j/h)$. The average distance between the electrodes is supposed to be $h$, and the distance between two lattice points is assumed to be $\mathfrak{d}$, which are connected by wires with self-capacitance $C_W\approx\epsilon_0 \mathfrak{d}$. The capacitance of the electrode is denoted by $C$. Then the hopping rate between the resonators can be calculated from $g=z_0^2U^2C^2C_W^2/[h^2(2C+C_W)^3]$. It is worth noting that the hopping rate $g$ can be modulated by varying the applied voltage $U$ [Fig.~\ref{Fig5}(b)]. The hopping rate is $g/2\pi\approx0.12~\rm{MHz}$ when $U=1~\rm{V}$, and it can reach $g/2\pi\approx12.33~\rm{MHz}$ for $U=10~\rm{V}$.

According to Eq.~(\ref{HTTSD}), the interaction between the oscillator array and the skyrmion qubits can be described by the Hamiltonian
\begin{equation}
    \begin{split}
        \hat{H}_{\rm{MTMS}}&=\sum_i \frac{\widetilde{\omega}_q}{2}\hat{\sigma}_z^i + \sum_i \omega_m \hat{a}_i^{\dagger}\hat{a}_i\\
        &+\sum_i \bar{\lambda}_{\rm{TS}}\left(\hat{a}_i+\hat{a}_i^{\dagger}\right)\hat{\sigma}_x^i\\ &+\sum_{i,j}g\left(\hat{a}_i+\hat{a}_i^{\dagger}\right)\left(\hat{a}_j+\hat{a}_j^{\dagger}\right) \\
        &-\sum_i \Omega_E\cos\left(2\omega_E t\right)\left(\hat{a}_i+\hat{a}_i^{\dagger}\right)^2.
    \end{split}
    \label{HMTMSOrigin}
\end{equation}
Here, we only consider the interactions between the nearest neighbors. By transforming to a rotating frame with frequency $\omega_E$ and employing the rotating wave approximation, Eq.~(\ref{HMTMSOrigin}) can be simplified to
\begin{equation}
    \begin{split}
        \hat{H}_{\rm{MTMS}}&=\sum_n \frac{\Delta_q}{2}\hat{\sigma}_z^n+\sum_n \Delta_m \hat{a}_n^{\dagger}\hat{a}_n \\ &+\sum_n\bar{\lambda}_{\rm{TS}}\left(\hat{a}_n\hat{\sigma}_+^n+H.c.\right) \\
        &+\sum_n g_n\left(\hat{a}_n^{\dagger}\hat{a}_{n+1}+H.c.\right) \\
        &-\sum_n \frac{1}{2}\Omega_E\left(\hat{a}_n^{\dagger}{^2}+\hat{a}_n^2\right),
    \end{split}
\end{equation}
\textcolor[rgb]{0,0,1}{where $g_n=z_0^2U_n^2C^2C_W^2/[h^2(2C+C_W)^3]$}. Utilizing the Bogoliubov transformation ($\hat{b}_n=\hat{a}_n\cosh r-\hat{a}_n^{\dagger}\sinh r$)~\cite{2016LemondeP1133811338,2019BurdP11631165,2021BurdP898902} and disregarding the anti-rotation term, we have Eq.~(\ref{HMTMSSqr}).

\section{\label{sec:appendixF}Tuning the hopping rate to implement a SSH chain}
The hopping rates $G_1$ and $G_2$ between phonons depend on the voltage $U$. One can realize the SSH model by adjusting the voltage $U$. At the beginning, it is assumed that the hopping rates between phonons are all equal, i.e., $G_1=G_2=G$, corresponding to a voltage of $U_0$. By adjusting the voltage $U$ we get $G_1=(1+\delta)G$ and $G_2=(1-\delta)G$, corresponding to the voltages $U_1$ and $U_2$, respectively. Using \textcolor[rgb]{0,0,1}{$g_n=z_0^2U_n^2C^2C_W^2/[h^2(2C+C_W)^3]$} and \textcolor[rgb]{0,0,1}{$G_n=g_n\exp(2r_n)/2$}, we get
\begin{equation}
    \begin{split}
        G_1=\frac{g_1}{2}\exp(2r_1)&=\frac{z_0^2U_1^2C^2C_W^2}{h^2(2C+C_W)^3}\frac{\exp(2r_1)}{2}\\
        &=(1+\delta)\frac{z_0^2U_0^2C^2C_W^2}{h^2(2C+C_W)^3}\frac{\exp(2r_0)}{2},\\
        G_2=\frac{g_2}{2}\exp(2r_2)&=\frac{z_0^2U_2^2C^2C_W^2}{h^2(2C+C_W)^3}\frac{\exp(2r_2)}{2}\\
        &=(1-\delta)\frac{z_0^2U_0^2C^2C_W^2}{h^2(2C+C_W)^3}\frac{\exp(2r_0)}{2}.
    \end{split}
\end{equation}
Then we can obtain the dependence of $\delta$ on squeezing parameter $r$ and voltage $U$ as
\begin{equation}
    \delta=1-\frac{U_2^2}{U_0^2}e^{2(r_2-r_0)}=\frac{U_1^2}{U_0^2}e^{2(r_1-r_0)}-1.
\end{equation}
That is, we can realize the modulation of $\delta$ by adjusting the voltage $U$ or the squeezing parameter $r$, and it satisfies the relation
\begin{equation}
    U_1^2e^{2(r_1-r_0)}+U_2^2e^{2(r_2-r_0)}=2U_0^2.
\end{equation}
In particular, when $r_1=r_2=r_0$, one can obtain $U_1^2+U_2^2=2U_0^2$. That is, we need to adjust the voltages to $U_1>U_0$ and $U_2<U_0$.

\section{\label{sec:appendixG}Skyrmion-skyrmion interactions mediated by chiral skyrmion-phonon bound states}
Here we analyze the skyrmion coupling to sublattice A as an example. By solving the secular equation $\hat{H}_{\rm{MTMS}}^{\rm{sq}}\vert \psi_{\rm{BS}}\rangle=E_{\rm{BS}}\vert \psi_{\rm{BS}}\rangle$ and the Fourier inverse transform~\cite{2022GongP5351753517}, the real-space phonon profile is given by
\begin{subequations}
    \begin{equation}
        C_{k,A}^A=\frac{\mathcal{G}C_eE_{\rm{BS}}}{2\pi}\int_{-\pi}^{\pi}dk \frac{e^{ikj}}{E_{\rm{BS}}^2-\Omega^2(k)},
        \label{CJ1}
    \end{equation}
    \begin{equation}
        C_{k,B}^A=\frac{\mathcal{G}C_e}{2\pi}\int_{-\pi}^{\pi}dk \frac{\Omega(k)e^{i[kj-\phi(k)]}}{E_{\rm{BS}}^2-\Omega^2(k)}.
        \label{CJ2}
    \end{equation}
    \label{CJ}
\end{subequations}
For the case $E_{\rm BS}=0$, calculating the integral~(\ref{CJ1}) we can get
For $E_{\rm BS}=0$, we can easily obtain $C_{k,A}^A=0$. Integral~(\ref{CJ2}) can also be computed to obtain
\begin{equation}
    C_{k,B}^A\left\{
    \begin{aligned}
        &\frac{\mathcal{G}C_e(-1)^j}{-G(1+\delta)}\left(\frac{1-\delta}{1+\delta}\right)^j,~j\geq0,\\
        &0,~j<0.
    \end{aligned}
    \right.
\end{equation}
This skyrmion-phonon bound state can be used to mediate chiral skyrmion-skyrmion interactions by exchanging virtual phonon. For $\delta>0$ and in the Markovian limit, the chiral skyrmion-skyrmion interaction can be written as~\cite{2022GongP5351753517,2022DongP2307723077,2019BelloP297297}
\begin{equation}
    \hat{H}_{\rm{SS}}=-\sum_{i<j}G_{i,j}\left(\hat{\sigma}_+^i\hat{\sigma}_-^j+H.c.\right),
\end{equation}
where the coupling strength is defined as $G_{i,j}^{AA/BB}=0$ and
\begin{equation}
    G_{i,j}^{AB}=\left\{
    \begin{split}
        &\frac{\mathcal{G}^2\left(-1\right)^{x_{ij}}}{G\left(1+\delta\right)}\left(\frac{1-\delta}{1+\delta}\right)^{x_{ij}}, x_{ij}\geq0, \\
        &0, x_{ij} < 0.
    \end{split}
    \right.
\end{equation}

%

\end{document}